\documentclass[prc,showpacs,amsmath,amssymb,11pt]{revtex4}
\usepackage{epsfig,graphicx}

\begin{document}
\title{Combined study of $\gamma p\rightarrow \eta p$ and
$\pi^-p\rightarrow \eta n$ in a chiral constituent quark approach}
\author{Jun He}
\email[]{junhe@impcas.ac.cn} \affiliation{Institute of Modern Physics, Chinese Academy of Sciences, Lanzhou 730000, P.R.China}
\author{B. Saghai}
\email[]{bijan.saghai@cea.fr} \affiliation{Institut de Recherche sur les lois Fondamentales de l'Univers,IRFU, CEA/Saclay, 
91191 Gif-sur-Yvette, France}
\date{\today}
\begin{abstract}
Within a chiral constituent quark model approach, $\eta$-meson production on the proton {\it via} electromagnetic and 
hadron probes are studied.
With few  parameters, differential cross-section and polarized beam asymmetry for $\gamma p~\to~\eta p$ and 
differential cross section for $\pi^- p\rightarrow \eta n$ processes are calculated and successfully compared with the data in 
the center-of-mass energy range from threshold up to 2 GeV.
The five known resonances $S_{11}(1535)$, $S_{11}(1650)$, $P_{13}(1720)$, $D_{13}(1520)$, and $F_{15}(1680)$ are found to be 
dominant in the reaction mechanisms in both channels.
Possible roles plaied by new resonances are also investigated and in the photoproduction channel, significant contribution 
from $S_{11}$- and $D_{15}$-resonances, 
with masses around 1715 MeV and 2090 MeV, respectively, are deduced.  
For the so-called missing resonances, no evidence is found within the investigated reactions.
The helicity amplitudes and decay widths of $N^* \to \pi N,~\eta N$ are also presented, and found consistent with the PDG values.
\end{abstract}
\pacs{  13.60.Le, 12.39.Fe, 12.39.Jh,  14.20.Gk }

\maketitle

\section{Introduction}\label{Sec:Intro}

Baryon spectroscopy is one of the major realms in deepening our understanding of QCD in the non-perturbative regime.
Properties of the nucleon and its resonances are extracted mainly through photo- and/or hadron-production of mesons 
off the nucleon.

In a recent paper~\cite{He2008a}, we investigated the $\gamma p\rightarrow \eta p$ process within a chiral constituent 
quark model and discussed the state-of-the art. 
In the present work, we extend that formalism to the $\pi^-p\rightarrow \eta n$ reaction and perform a combined analysis
of both channels.

For the photoproduction process, a healthy amount of data has been 
released in recent years for both differential cross section~\cite{Dugger2002,Crede2005,Nakabayashi2006,Bartalini2007}, 
and polarized beam asymmetry~\cite{Elsner2007,Bartalini2007}. The situation is very different for the 
$\pi^{-}p \rightarrow \eta n$ reaction. Actually, the data come mainly from measurements performed in 
70's~\cite{Deinet1969,Richards1970,Debenham1975,Brown1979,Bulos1969,Feltesse1975} and suffer from some 
inconsistencies~\cite{Clajus1992}.
A recent experiment, performed at BNL using the Crystal Ball spectrometer~\cite{Prakhov2005}, offers a high quality data set,
though limited to the close to threshold kinematics. Consequently, a combined data base embodying experimental results for both
electromagnetic and strong channels turns out to be highly heterogeneous. In spite of that uncomfortable situation, recent
intensive theoretical investigations interpreting both channels within a single approach has proven to be  fruitful
in revealing various aspects of the relevant reaction mechanisms, as discussed, e.g., in Refs.~\cite{He2008a,Durand2008}. 

In the photoproduction sector, a significant progress has been performed in recent years within coupled-channels
formalisms~\cite{Chiang2003a,Feuster1999,Sarantsev2005,Arndt2007} allowing to investigate a large number of 
intermediate and/or final meson-baryon (MB) states: 
$\gamma N \to MB$, with $MB \equiv \pi N, \eta N, \rho N, \sigma N, \pi \Delta, K \Lambda, K \Sigma$.
Those approaches have been reviewed in our recent paper~\cite{He2008a}.
Also advanced coupled-channels approaches are being developed~\cite{Chiang2004,JLMS2007,Durand2008,Durand2009a} 
for the strong channels:
$\pi N \to MB$.
However, fewer studies embody both electromagnetic and strong production processes.
Moreover, those works are based on the effective Lagrangian approaches (ELA), where meson-baryon degrees
of freedom are implemented, (see e.g. Refs.~\cite{Feuster1999,Penner2002,Anisovich2005,Sarantsev2005}). 
Investigations based on subnucleonic degrees of freedom, via
constituent quark models (CQM) have been successful~\cite{He2008a,Saghai2001,Zhao2002,Julia-Diaz2006,Saghai2007,He2008} 
in the interpretation of photoproduction data on the proton, 
namely, $\gamma p \to \pi N,~\eta p,~K \Lambda$, and a recent work~\cite{Zhong2007} has considered the 
$\pi^{-}p \rightarrow \eta n$ reaction.

At the present stage, the ELA and the CQM approaches are complementary. However, the QCD-inspired CQM developments deal 
on the one hand with more fundamental degrees of freedom and on the other hand require a much smaller number of
adjustable parameters while fitting the data. This latter feature allows including a large number of resonances in the model
search with still a reasonable number of free parameters. Hence, this approach turns out to be suitable in searching
for the so-called missing and/or new resonances
~\cite{Li1996,Kelkar1997,Batinic1997,Saghai2001,Giannini2001,Chen2003,Chiang2003,Tryasuchev2004,Mart2004,Sarantsev2005,Ablikim2006,Fang2006,Julia-Diaz2006}.

The present work is hence a step in a combined study of both electromagnetic and strong $\eta$-production processes
within a unified chiral constituent quark ($\chi QM$) formalism.

The paper is organized as follows.  
In Section~\ref{Sec:Theo}, the theoretical content of our $\chi QM$ approach is presented. 
The fitting procedure and numerical results for differential cross-section, polarized beam asymmetry, 
helicity amplitudes, and partial decay widths are reported and discussed in section~\ref{Sec:Res}, 
where possible roles played by ``missing'' and new resonances are also examined. 
Summary and conclusion are given in section~\ref{Sec:Conclu}.

\section{Theoretical Frame}\label{Sec:Theo}

To investigate hadrons and their resonances, various formalisms embodying the subnucleonic degrees of freedom
are being developed.
Lattice QCD, based on the fundamental theory of strong interactions, is expected to establish the properties of
hadrons, but there are still great technical difficulties when applied to resonances, see e.g.~\cite{Basak2007,Lang2008}. 
The QCD sum rule approach is also applied to the resonance region, though limited
to the low mass ones, such as $\Delta(1232)$ and
$S_{11}(1535)$~\cite{Lee1997,Jido1998,Zhu1999a,Lee2007,Lee2002}. That technique faces difficulties to control the
uncertainties in handling phenomenological parameters.
The most efficiently used approach to study the baryon resonance is the constituent
quark model, which provided the first clear evidence of the underlying $SU(6)\otimes O(3)$ 
structure of the baryon spectrum~\cite{Copley1969}. Subsequent studies have been concentrating mainly on the 
transition amplitudes and the baryon mass spectrum, achieving well known 
successes~\cite{Capstick2000,Copley1969,Feynman1971,Koniuk1980,Capstick1992}, but do not investigate reaction mechanisms.

To connect the constituent quark model to the reaction mechanisms of specific processes, a comprehensive and unified approach 
to the pseudoscalar mesons photoproduction, based on the low energy QCD
Lagrangian~\cite{Manohar1984}, is developed ~\cite{Li1997}, and applied to some processes, including
$\gamma p\rightarrow \eta p$~\cite{Li1998,Saghai2001,He2008} and $\pi^-p\rightarrow \eta n$
~\cite{Zhong2007,He2007}.

In this section we recall briefly the content of the chiral constituent quark approach~\cite{He2008a} and extend it to the 
$\eta$ hadron-production process. As in Ref.~\cite{Li1997} we start from an effective chiral
Lagrangian~\cite{Manohar1984},

\begin{eqnarray} \label{lg}
\mathcal{L}=\bar{\psi}[\gamma_{\mu}(i\partial^{\mu}+V^{\mu}+\gamma_5A^{\mu})-m]\psi
+\cdot\cdot\cdot,
\end{eqnarray}
\noindent where vector ($V^{\mu}$) and axial ($A^{\mu}$) currents
read,
\begin{eqnarray}
V^{\mu} =
 \frac{1}{2}(\xi\partial^{\mu}\xi^{\dag}+\xi^{\dag}\partial^{\mu}\xi)~,~A^{\mu}=
 \frac{1}{2i}(\xi\partial^{\mu}\xi^{\dag}-\xi^{\dag}\partial^{\mu}\xi),
\end{eqnarray}
with $ \xi=\exp{(i \phi_m/f_m)}$ and $f_m$ the meson decay constant.
$\psi$ and $\phi_m$ are the quark and meson fields, respectively.

In this paper we focus on the resonance contributions, for which the
amplitudes can be written as,
\begin{equation}\label{42}
{\cal M}_{N^*}=\frac {2M_{N^*}}{s-M_{N^*}^2-iM_{N^*}\Gamma({\bf q})}e^{-\frac {{\bf
k}^2+{\bf q}^2}{6\alpha^2}} {\cal O}_{N^*},
\end{equation}
where $\sqrt {s} \equiv W$ is the total
energy of the system, and ${\cal O}_{N^*}$ is determined by the
structure of each resonance. $\Gamma({\bf q})$ in Eq. (\ref{42}) is
the total width of the resonance, and a function of the final state
momentum ${\bf q}$.  

The transition amplitude for the $n^{th}$ harmonic-oscillator shell is
\begin{eqnarray}\label{On}
{\cal O}_{n}={\cal O}_n^2 +{\cal O}_n^3.
\end{eqnarray}

The first (second) term represents the process in which the incoming
photon and outgoing meson, are absorbed and emitted by the
same (different) quark.

We use the standard multipole expansion of the
CGLN amplitudes~\cite{Chew1957} to obtain the partial wave
amplitudes of resonance $f_{2I, 2l\pm1}$. Then the transition
amplitudes for peseudoscalar meson production through photon and meson baryon scattering takes, respectively, 
the following form:
\begin{eqnarray}
\label{63} {\cal O}^\gamma_{N^*}&=&if_{1l\pm}  {\bf \sigma} \cdot {\bf
\epsilon}+ f_{2l\pm} {\bf \sigma} \cdot {\bf \hat{q}} {\bf \sigma}
\cdot ({\bf \hat{k}} \times {\bf \epsilon})+ if_{3l \pm} {\bf
\sigma} \cdot {\bf \hat{k}} {\bf \hat{q}} \cdot {\bf \epsilon} +
if_{4l\pm}  {\bf \sigma} \cdot {\bf \hat{q}}{\bf \epsilon}\cdot {\bf
\hat{q}},\nonumber\\
{\cal O}^m_{N^*}&=&f_{1l\pm}+{\bf
\sigma}\cdot\hat{\bf q}{\bf \sigma}\cdot\hat{\bf k}f_{2l\pm}.
\end{eqnarray}

In Ref.~\cite{Li1997}, the partial decay amplitudes are used to
separate the contribution of the state with the same orbital angular
momentum $L$. As we found in Ref.~\cite{He2008a}, with the helicity amplitudes of photon
transition and meson decay, we can directly obtain the CGLN amplitudes for each resonance in terms of 
Legendre polynomials derivatives.
Analogously, the partial wave amplitudes for pseudoscalar meson-baryon scattering are
\begin{eqnarray}f_1&=&\sum_{l=0}^\infty[f_{l+}P'_{l+1}-f_{l-}P'_{l-1}],
\nonumber\\
f_2&=&\sum_{l=0}^\infty[f_{l-} -f_{l+}]P'_{l}.
\end{eqnarray}

We can connect the helicity amplitudes with the multipole coefficients as in the case of photoproduction process
\begin{eqnarray}
  f_{l\pm}&=& \mp A_{l\pm}\simeq\frac{1}{2}\epsilon \left(
  \frac{\Gamma_{m_i} \Gamma_{m_j}}{kq} \right)^{1/2}C^I_{m_i N}C^I_{m_j N} \;
= \frac{1}{2\pi(2J+1) }
[\frac{ E_{N_i}E_{N_i}}{
M^2_{N^*}}]^{1/2}A^{m_i}_{1/2}A^{m_j}_{1/2};,
\end{eqnarray}
where $C^I_{m N}$ represents the Clebsch-Gordan coefficients
related to the isospin coupling in the incoming or outgoing channel with
$m_i$ and $m_j$ the incoming and outgoing mesons (respectively $\pi$ and $\eta$ in this work).

In our approach, the photoexcitation helicity amplitudes
$A_\lambda^\gamma$, as well as the strong decay
amplitudes $A^m_\nu$, are related to the matrix elements of interaction Hamiltonian~\cite{Copley1969} as following,
\begin{eqnarray}A_\lambda  &=&\sqrt{\frac{2\pi}{k}}\langle
{N^*};J\lambda|H_{e}|N;\frac{1}{2}\lambda-1\rangle,  \\
A^m_\nu&=&\langle N;\frac{1}{2}\nu|H_{m}|{N^*};J\nu\rangle.
\end{eqnarray}

The amplitudes in Ref.~\cite{Li1997} are derived under the
$SU(6)\otimes O(3)$ symmetry. However, for physical states that
symmetry is broken. An example is the violation of the Moorhouse
rule~\cite{Moorhouse1966}. In Ref.~\cite{Li1998}, a set of parameters
$C_{{N^*}}$ were hence introduced to take into account the breaking
of that symmetry, {\it via} following substitution:
\begin{eqnarray}\label{eq:AR}
{\mathcal O}_{N^*} \to C_{N^*} {\mathcal O}_{N^*} .
\end{eqnarray}
In Refs.~\cite{Li1998,Saghai2001}, those parameters were allowed to
vary around their $SU(6)\otimes O(3)$ values ($\vert C_{N^*} \vert$
= 0 or 1). In this work, instead of using those adjustable
parameters, we introduce the breakdown of that symmetry through the
configuration mixings of baryons wave functions as we have done in Ref.~\cite{He2008a}.
To achieve that improvement, we adopted the one-gluon-exchange (OGE)
model~\cite{Isgur1978,Isgur1978a,Isgur1979}, which has been
successfully used to study the helicity amplitudes and decay
widths of resonances~\cite{Koniuk1980}.

%

\section{Results and discussion}\label{Sec:Res}

The most important and interesting nucleon resonances, such as $S_{11}(1535)$, are in the mass region
lower than 2 GeV, that is, the $n$=1,2 shell states in the constituent quark
model~\cite{Isgur1977,Isgur1978}. 
In this region plentiful recent data are expected to give more reliable information about the 
internal structure and properties of baryon resonances. 
Hence, in the present work we investigate the reactions $\gamma p \to \eta p$
and $\pi^-p\rightarrow \eta n$, focusing on the range of centre-of-mass total
energy from threshold up to $W~\approx$ 2 GeV,  in order to interpret a
large amount of high quality data released from various facilities.

%
\subsection{Fitting procedure}\label{Sec:Fit}
Using the CERN MINUIT code, we have fitted simultaneously the following data sets and the PDG values:

\begin{itemize}
\item {\bf Spectrum of known resonances:}
    \subitem {\it Known resonances:}
 we use as input the PDG values~\cite{Yao2006} for masses and widths, with the
uncertainties reported there plus an additional theoretical
uncertainty of 15 MeV, as in Ref.~\cite{Capstick1992}, in order to
avoid overemphasis of the resonances with small errors. The data
base contains
 all 12 known nucleon resonances
 as in PDG, with $M~\le$~2 GeV, namely,

{\boldmath$ n$}{\bf =1:} $S_{11}(1535)$, $S_{11}(1650)$,
$D_{13}(1520)$, $D_{13}(1700)$, and $D_{15}(1675)$;

{\boldmath$ n$}{\bf =2:} $P_{11}(1440)$, $P_{11}(1710)$,
 $P_{13}(1720)$, $P_{13}(1900)$,
$F_{15}(1680)$, $F_{15}(2000)$, and $F_{17}(1990)$.

Besides the above isospin-1/2 resonances, we fitted also the mass of
$\Delta$(1232) resonance. However, spin-3/2 resonances do not
intervene in the $\eta$ photoproduction. 
Concerning the resonances for which uncertainties are not given in PDG, we use 50 MeV.
 \subitem {\it Additional resonance:} Resonances with masses above $M~\approx$~2 GeV,
treated as degenerate, are simulated by a single resonance, for
which are left as adjustable parameters the mass, the width, and the
symmetry breaking coefficient.
\item {\bf  Observables for $\gamma p\rightarrow \eta p$:}
\subitem {\it Differential cross-section:} Data base includes 1220 data points
for 1.49 $\lesssim W \le$ 1.99 GeV, coming from the following labs:
MAMI~\cite{Krusche1995} (100 points), CLAS~\cite{Dugger2002} (142 points),
ELSA~\cite{Crede2005} (311 points), LNS~\cite{Nakabayashi2006} (180 points), and
GRAAL~\cite{Bartalini2007} (487 points). Only statistical uncertainties are used.

 \subitem {\it Polarized beam asymmetry:} 184 data points
for 1.49 $\lesssim W \le$ 1.92 GeV,
 from GRAAL~\cite{Bartalini2007} (150 points) and ELSA~\cite{Elsner2007} (34 points). Only
statistical uncertainties are used.

 \subitem {\it Target asymmetry:}
 The target asymmetry ($T$) data~\cite{Bock1998} are not
included in our data base. Actually, those 50 data points bear too
large uncertainties to put significant constraints on the
parameters~\cite{He2008}.
\item {\bf  Observables for $\pi^-p\rightarrow \eta n$:}
\subitem {\it Differential cross-section:} Data base includes 354 data points,
for 1.49 $\lesssim~W~\le$ 1.99 GeV, coming from: Deinet~\cite{Deinet1969} (80 points),
Richards~\cite{Richards1970} (64 points), Debenham~\cite{Debenham1975} (24 points), Brown~\cite{Brown1979} (102 points),
Prakhov~\cite{Prakhov2005} (84 points).  Uncertainties are treated as in Ref.~\cite{Durand2008}
\end{itemize}
As already mentioned, for the $\pi^-p\to \eta n$ process, the data set is composed
mainly of old data, plus those released recently by Prakhov {\it et al}~\cite{Prakhov2005}.
Models constructed~\cite{Zhong2007,He2007,Durand2008} using those experimental results encountered some
difficulties in reproducing especially the two lowest energy data sets.
Those features deserve a few comments. 
The Prakhov {\it et al.}~\cite{Prakhov2005} data set consists of differential cross section in nine incident 
pion momentum bins in the range $P_\pi$ = 687 to 747 MeV/c, corresponding to the center-of-mass energy range 
$W$ = 1.49 to 1.52 GeV.
It is interesting to notice that the reported total cross section increases by almost one order of magnitudes going from the
lowest to the highest pion momentum. In order to attenuate the undesirable effects of such sharp variations, we introduce
an energy dependent term in the denominator of the $\chi^2$ expression used in the minimization procedure, namely,
\begin{eqnarray}\label{eq:chi2} 
\chi^2=\sum\frac{(V_{ex}-V_{th})^2}{(\delta V_{ex})^2+(V'_{th} \Delta E_{ex})^2} 
\end{eqnarray}
Here $V_{ex}$, $V_{th}$, and $\delta V_{ex}$ are the standard $\chi^2$ quantities. The additional term is a product
of the derivative of the observable with respect to energy ($V'_{th}$), and the experimental energy bin ($\Delta E_{ex}$).
Notice that the data are reported for central values of $P_\pi \pm \Delta P_\pi$, with $\Delta P_\pi$ = 3 to 7 MeV/c.
We will come back to this point.

In summary, 1783 experimental values are fitted. To do so, we have a total of 21 free parameters, not all
of them adjusted on all the data sets, as explained below.

In Table~\ref{Tab: Para} we report the list of adjustable parameters and their extracted values.

\begin{table*}[ht!]
\caption{Adjustable parameters with extracted values and $\chi^2$, where
$m_q$, $\alpha$,  $\Omega$, $\Delta$, $M$, and $\Gamma$ are in
MeV.}
\renewcommand\tabcolsep{0.5cm}
\begin{tabular}{ll|ccc}  \hline\hline
   & Parameter  & Model $B$ in Ref.~\cite{He2008a} &This work     \\ \hline
  & $g_{\eta NN}$              & 0.449             &  0.376       \\ \hline
  & $m_q$                      & 304               &  312       \\
  & $\alpha$                   & 285               &  348       \\
  & $\alpha_s$                 & 1.98              &  1.96        \\
  & $\Omega$                   & 442               &  437         \\
  & $\Delta$                   & 460               &  460         \\ \hline
 HM $N^*$:    &$M$             & 2129              & 2165         \\
              &$\Gamma$        &   80              & 80           \\
              &$C^\gamma_{N^*}$& -0.70             &-0.84         \\
              &$C^\pi_{N^*}$   & $--$              & 0.005         \\\hline
$P_{13}(1720)$:
  & $C^\gamma_{P_{13}(1720)}$  & 0.40              &  0.37        \\
  & $C^\pi_{P_{13}(1720)}$     & $--$              & -0.89        \\ \hline
New $S_{11}$: &$M^\gamma$      & 1717              & 1715         \\
              &$\Gamma^\gamma$ &  217              & 207          \\
              &$C^\gamma_{N^*}$& 0.59              & 0.51         \\
New $D_{13}$: &$M^\gamma$      & 1943              & 1918         \\
              &$\Gamma^\gamma$ &  139              & 151          \\
              &$C_{N^*}^\gamma$& -0.19             & -0.19        \\
New $D_{15}$: &$M^\gamma$      & 2090              & 2090         \\
              &$\Gamma^\gamma$ &  328              & 345        \\
              &$C_{N^*}^\gamma$& 2.89              & 2.85         \\
\hline \hline
$\sum\chi^2_{dp}/N_{dp}$:
              &$\chi^2$ for total           & 3273/1418=2.31       &3627/1772=2.05          \\
&$\chi^2_\gamma$ for $\gamma p \to \eta p$         & 3243/1404=2.31       &3187/1404=2.27          \\
&$\chi^2_\pi$ for $\pi^-p\rightarrow \eta n$    & $--$                 &408/354=1.15         \\
 &   Spectrum                               & 30/14=2.14                &32/14=2.29             \\
\hline \hline\end{tabular} \label{Tab: Para}

\end{table*}

Two of the parameters, namely, the non-strange quarks average mass ($m_q$) and the harmonic oscillator strength
($\alpha$) are involved in fitting both mass spectrum and $\eta$-production
data. The QCD coupling constant ($\alpha_s$) and the confinement constants
($\Omega$ and $\Delta$), intervene only  in fitting the $\eta$-production data {\it via} the 
configuration mixing mechanism.

In Table~\ref{Tab: Para} the extracted values within the present work are compared to those reported in 
our previous paper.
The only significant variation concerns the harmonic oscillator strength ($\alpha$), which is lowered by
some 20\%, due to inclusion of the strong channel.
The quark mass is very close to the commonly used values, roughly one third of the nucleon mass. 
For the other parameters, the extracted values here come out close to those used by 
Isgur and Capstick~\cite{Isgur1979,Capstick2000}: $E_0$~=~1150 MeV, $\Omega$ $\approx$ 440 MeV, and
$\Delta$ $\approx$ 440 MeV. 
For the parameters $\alpha_s$, $\alpha$, and $m_q$ Isgur and Capstick introduce 
$\delta$ = $(4 \alpha_s \alpha) / (3 \sqrt{2 \pi} m_{q}^2)$, for which they get $\approx$ 300 MeV. 
Our model gives $\delta$ $\approx$ 262 MeV.

Among the remaining 16 adjustable parameters, 9 of them (related to the new resonances) are extracted by fitting 
the photoproduction data, and the additional 7 parameters are determined by fitting
data for both channels.
With respect to the $\eta$-nucleon coupling constant $g_{\eta NN}$, our result favors a rather small
coupling around $g_{\eta NN} = 0.376$, which is compatible with those deduced
from fitting only the $\eta$ photoproduction~\cite{Li1998,Saghai2001}.
Comparable values for the coupling are also reported in Refs.~\cite{Tiator1994,Kirchbach1996,Zhu2000,Stoks1999}.

The parameter $C_{P_{13}(1720)}$ is the strength of the ${P_{13}(1720)}$ resonance, that we had to leave as
a free parameter in order to avoid its too large contribution resulting from direct calculation, as
discussed in Ref.~\cite{He2008a}. The value of that parameter for the photoproduction reaction is close
to that in our previous paper. For the strong channel, $C^\pi_{P_{13}(1720)}$ turns out to be larger in magnitude than
the of photoproduction ($C^\gamma_{P_{13}(1720)}$).

The higher mass resonance (HM $N^*$) treatment requires four adjustable parameters: $M$, $\Gamma$,
$C^\gamma_{N^*}$, and $C^\pi_{N^*}$, which are determined by fitting the $\eta$-production data.   
Here, different strengths ($C^\gamma_{N^*}$, and $C^\pi_{N^*}$) for higher mass resonances and ${P_{13}(1720)}$ are 
used because two processes have different initial states.
Notice that in fitting the $\eta$-production data, we use the PDG~\cite{Yao2006} values for
masses and widths of the known resonances.

In recent years, several
authors~\cite{Sarantsev2005,Saghai2001,Li1996,Batinic1997,Giannini2001,Ablikim2006,
Fang2006,Chen2003,Chiang2003,Tryasuchev2004,Mart2004,Kelkar1997,Julia-Diaz2006}
have put forward need for new resonances in interpreting various observables, with extracted masses 
roughly between 1.7 and 2.1 GeV. 
We have hence, investigated possible contributions from three of 
them: $S_{11}$, $D_{13}$, and $D_{15}$. 
For each of those new resonances we introduce then three additional adjustable parameters per resonance: 
mass~($M$), width~($\Gamma$), and symmetry breaking coefficient~($C_{N^*}$). 
For the three new resonances, we follow the method in Ref.~\cite{Li1998}
{\it via} Eq.~(\ref{eq:AR}). 
The extracted Wigner masses and widths, as well as the strengths for those resonances are given in 
Table~\ref{Tab: Para}.
The results are close to ones in Ref.~\cite{He2008a}. 

For the process $\pi^-p\rightarrow \eta n$, given the state of the data base, the determination
of the reaction mechanism is less reliable than for the photoproduction. 
Consequently, search for signals from unknown resonances in that strong channel would be superfluous.
Nevertheless, we looked at possible contributions from those three new resonances and found their
contributions negligible. Accordingly, for the strong channel we deal only with the known resonances.

As shown in Table \ref{Tab: Para}, the $\chi^2$ for both processes
is 2.05, with 2.27 for the $\eta$ photoproduction and 1.15 for
$\pi^-p\rightarrow\eta n$. So, within our model, the data are well enough
reproduced. With respect to the latter channel,
if we do not consider the uncertainty for the energy (Eq.~(\ref{eq:chi2})), and use the same definition for
$\chi^2$  as the EBAC collaboration~\cite{Durand2008}, we obtain $\chi^2=1.99$ for
$\pi^-p\rightarrow\eta n$, which is close to its EBAC value, 1.94.

To end this section, we examine the role played by each of the 12 known, 3 new, and one heavy mass resonances.
To that end, we have switched off resonances one by one. 
The results are reported in Table~\ref{Tab:chi2}. 
For each case, two numbers are given corresponding to two sets of $\chi^2$s: 
i) without further minimization and ii) after minimization (in brackets).  
That Table embodies results for 7 resonances. 
For the remaining 9 resonances the variations of $\chi^2$ were found negligible and hence, 
not shown in the Table.
However, in some regions in the phase space, few of those resonances play significant roles and
we will emphasize those features in the relevant sections.

\begin{table*}[ht!]
\caption{ $\chi^2$s after turning off the corresponding resonance contribution,
without [with] further minimizations, including partial $\chi^2$s for the $\gamma p \to \eta p$ and 
$\pi^-p\rightarrow \eta n$ processes.}
\renewcommand\tabcolsep{0.16cm}
\begin{tabular}{ lllllllll }  \hline\hline
                 & $S_{11}$(1535)&$S_{11}$(1650)& $P_{13}$(1720)& $D_{13}$(1520)&$F_{15}$(1680)& New $S_{11}$ & New $D_{15}$ \\\hline
$\chi^2$         & 136 [80 ]     & 12.3 [2.36 ] & 3.49 [2.90]   & ~9.60 [5.27]   & 4.33 [3.25]  & ~9.63 [4.44]  & 3.37 [2.34]  \\
$\chi^2_\gamma$  & 160 [85 ]     & 14.2 [2.60 ] & 3.55 [2.77]   & 10.04 [5.79]   & 4.44 [3.59]  & 11.84 [5.22]  & 3.93 [2.60]  \\
$\chi^2_\pi$     & ~48 [65 ]     & ~4.9 [1.40 ] & 3.31 [3.38]   & ~8.17 [3.32]   & 3.98 [1.91]  & ~1.16 [1.39]  & 1.16 [1.28]   \\
\hline \hline
\end{tabular}
\label{Tab:chi2}
\end{table*}

For the known resonances in PDG, as expected, the most important role is played by the $S_{11}$(1535). The effects
of $S_{11}$(1650), $D_{13}$(1520), and to a less extent those of $F_{15}$(1680) and $P_{13}$(1720), turn out to be (very) significant. 
In addition to those known resonances, a new $S_{11}$ appears to be strongly needed by the photoproduction data. 
We also investigated possible contributions from the missing resonances (see next section) following the above procedure 
and found no significant effects. 

In the following , we will first present our results for the baryon spectrum. 
Then we will move to the observables for the $\gamma p\rightarrow \eta p$  and $\pi^- p\rightarrow \eta n$ processes,
and compare our model with the data. To get better insights to the reaction mechanism, we will also report on our 
results obtained by turning off the resonances which have significant effects on $\chi^2$ for both
processes studied here without further minimizations.

\subsection{Baryon spectrum}\label{Sec:BS}

The results of baryon spectrum extracted from the present work are
reported in Tables~\ref{Tab: Spectrum}. 
Our results are in good agreement with those obtained by Isgur and
Karl~\cite{Isgur1978a,Isgur1979}, and except for the $S_{11}$(1535)
and $D_{13}$(1520), fall in the ranges estimated by
PDG~\cite{Yao2006}. The additional "missing" resonances generated by
the OGE model, are also shown in Table~\ref{Tab: Spectrum}. The extracted
masses are compatible with those reported by Isgur and Karl~\cite{Isgur1978a,Isgur1979}.

\begin{table*}[ht!]
\caption{Extracted masses for known PDG and the so-called ``missing''
resonances compared with the values by Isgur {\it et al.}~\cite{Isgur1978a,Isgur1979} and PDG\cite{Yao2006}.
All are in the unit MeV.}
\renewcommand\tabcolsep{0.2cm}
\begin{tabular}{ l|cccccc }  \hline\hline
 PDG resonances                    & $S_{11}$(1535)&$S_{11}$(1650)     & $P_{11}$(1440)   &$P_{11}$(1710)&$P_{13}$(1720)&$P_{13}$(1900)\\ \hline
$M^{OGE}$ in this work             & 1471          & 1617              & 1423             & 1720         & 1712         & 1847 \\
$M^{OGE}$ in Refs.~\cite{Isgur1978a,Isgur1979}  & 1490          & 1655              & 1405             & 1705         & 1710         & 1870 \\
$M^{PDG}$ ~\cite{Yao2006}          &$1535 \pm 10$  &$1655^{+15}_{-10}$ &$1440^{+30}_{-20}$&$1710\pm 30$  &$1720^{+30}_{-20}$& $1900$ \\\hline

PDG resonances                     & $D_{13}$(1520)&$D_{13}$(1700)     &$D_{15}$(1675)    &$F_{15}$(1680)&$F_{15}$(2000)&$F_{17}$(1990)\\ \hline
$M^{OGE}$ in this work             & 1509          & 1697              & 1629             & 1717         & 2002         & 1939 \\
$M^{OGE}$ in Refs.~\cite{Isgur1978a,Isgur1979}  & 1535          & 1745              & 1670             & 1715         & 2025         & 1955 \\
$M^{PDG}$ ~\cite{Yao2006}          &$1520 \pm 5$   & $1700 \pm 50$     & $1675\pm 5  $    & $1685 \pm 5 $& $2000$       & $1990$ \\
\hline \hline
  ``missing resonances''             & $P_{11}$      &$P_{11}$           & $P_{13}$         &$P_{13}$      &$P_{13}$      &$F_{15}$\\ \hline
$M^{OGE}$ in this work             & 1893          & 2044              & 1936             & 1959         & 2041         & 1937 \\
Ref.~\cite{Isgur1979}              & 1890          & 2055              & 1955             & 1980         & 2060         & 1955 \\
\hline \hline
\end{tabular}
\label{Tab: Spectrum}
\end{table*}

\subsection{Observables for $\gamma p\rightarrow \eta p$}\label{Sec:Diff}

This section is devoted to the results for differential cross sections ($d\sigma/d\Omega$) and polarized beam asymmetries
($\Sigma$) for $\gamma p\rightarrow \eta p$, as in our previous work \cite{He2008a}, as well as those for total
cross section. 

\begin{figure}[ht!]
\includegraphics[bb=15 20 460 260 ,scale=1.15]{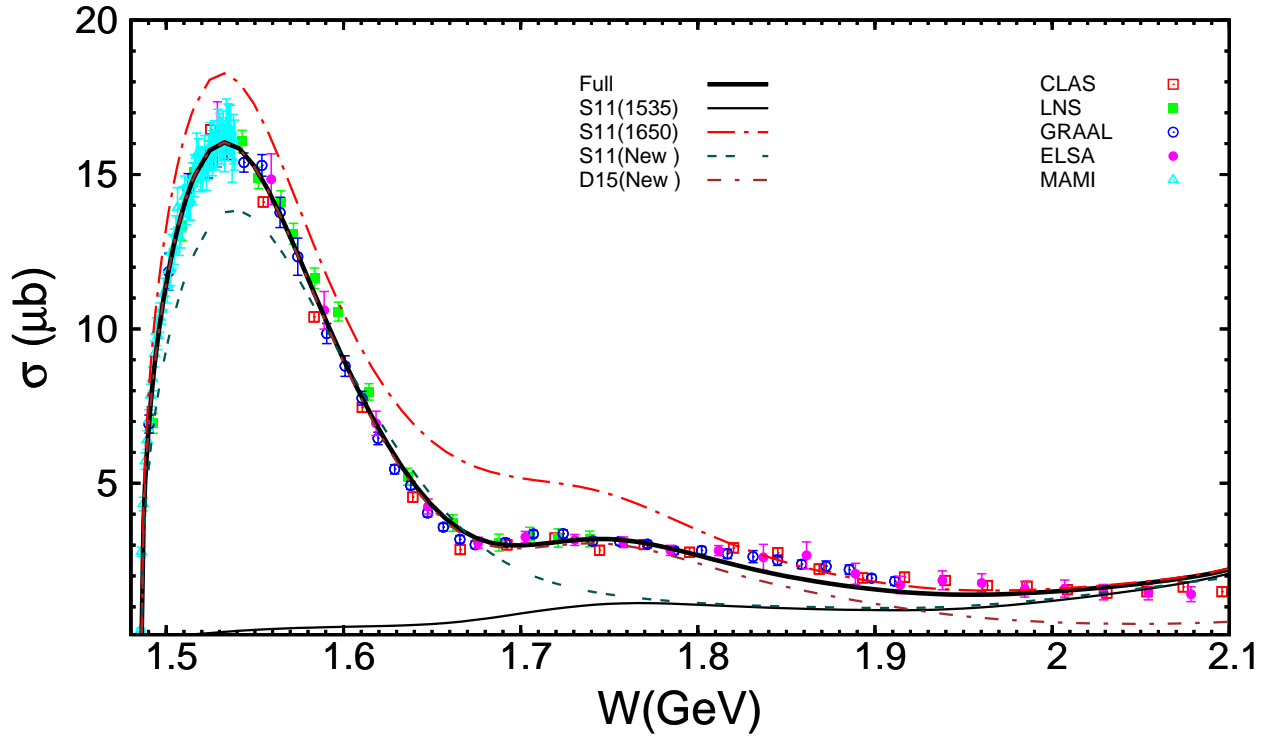}
\caption{(Color online) Total cross section for $\gamma p\rightarrow \eta p$ as a 
function of total centre-of-mass energy $W$. 
The curves are: full model (thick full), and turning off the following resonances: $S_{11}(1535)$ (thin full),
$S_{11}(1650)$ (long dash-dotted), new $S_{11}$ (dash-dashed), and
new $D_{15}$ (short dash-dotted). Data are from
CLAS~\cite{Dugger2002}, LNS~\cite{Nakabayashi2006}, GRAAL~\cite{Bartalini2007}, ELSA~\cite{Crede2005},
and MAMI~\cite{Krusche1995}. 
\label{Fig:gpept}}
\end{figure}

First, to give an overall picture of various features of our model, we present the results for total cross section 
and compare them to the data (Fig.~\ref{Fig:gpept}). Notice that the data for total cross section are not
used in minimization.
%
%

Our full model gives a reasonable account of the total cross section behavior from threshold up to 2 GeV with 
a small discrepancy around $W$=1.9 GeV. In Fig.~\ref{Fig:gpept} we also show results obtained by turning
off the most significant resonances, without further minimizations.

Switching off the $S_{11}(1535)$, the close threshold cross section decreases by more than two 
orders of magnitude. At energies far above threshold, the absence of that resonance shows still
non-negligible effects.
The contribution of the second $S_{11}$ resonance, $S_{11}(1650)$, is visible
from about 1.55 GeV upto 1.75 GeV. As reported in Table~\ref{Tab:chi2}, turning off that resonance
increases the $\chi^2$ from 2.05 to 12.3, without further minimization, and to 2.36 after minimization.
The significant discrepancy between the two $\chi^2$s can be understood by the fact that in the same
mass region, there are two other relevant resonances (new $S_{11}$ and $P_{13}(1720)$), and by redoing minimizations
their relative strengths get new values. Such a "compensating" mechanism shows up also in effective
Lagrangian based models through the extracted $\eta N N^*$ couplings.
The new $S_{11}$ resonance turns out playing a significant role roughly between 1.7 and 1.8 GeV.
Finally, a new $D_{15}$ appears, affecting the highest energy region  investigated here.

%
%
\begin{figure}[ht!]
\includegraphics[bb=180 50 470 460 ,scale=0.83]{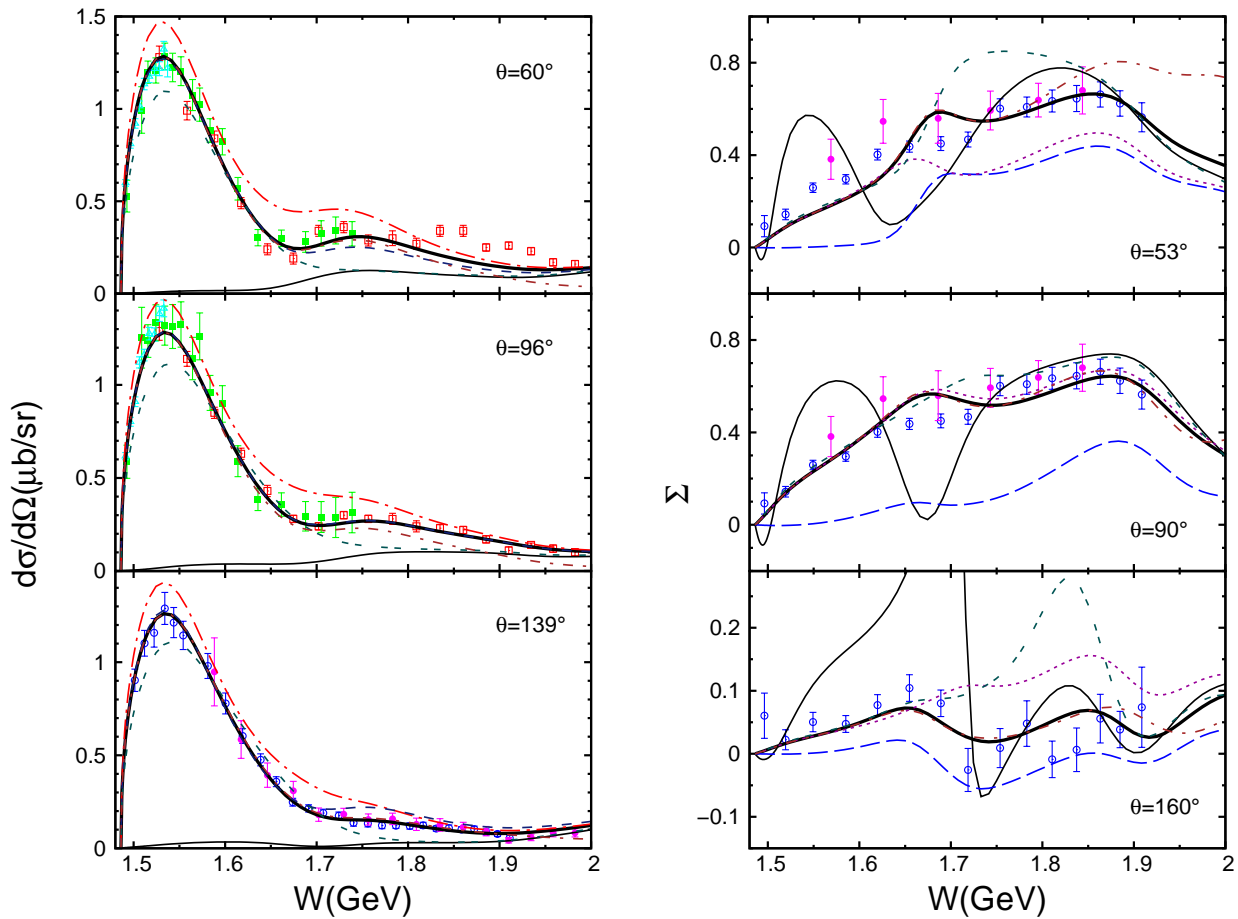}\
\caption{(Color online) Differential cross section (left panel) and  polarized
beam asymmetry (right panel), for $\gamma p\rightarrow \eta p$, as a function of $W$ at three angles. 
The curves are: {\bf both panels}: full model (thick full), and turning off $S_{11}(1535)$ (thin full),
new $S_{11}$ (dash-dashed), and new $D_{15}$ (short dash-dotted);
{\bf left panel}: switching off $S_{11}(1650)$ (long dash-doted), $P_{13}(1720)$ (dashed); 
{\bf rigth panel}: turning off $D_{13}(1520)$(dashed),
$F_{15}(1680)$ (dotted). 
Data are as in Fig.~\ref{Fig:gpept}.
\label{Fig: gpep}}
\end{figure}

Excitation functions for differential cross sections and polarized beam asymmetries are presented in 
Fig~\ref{Fig: gpep}, left and right panels, respectively. 
For differential cross section, model / data comparisons lead to similar conclusions as in the
case of total cross section, with respect to the model ingredients. 
Moreover, the underestimate of data around 1.85 to 1.95 GeV  turns out
to happen at forward angles.

Results for the polarized beam asymmetries are shown in the right panel in Fig~\ref{Fig: gpep}.
Here also the $S_{11}(1535)$ is still the most dominant ingredient upto roughly 
1.75 GeV.
In contrast to the cross section reaction mechanism, we observe significant contributions from two known
resonances. Actually, the $D_{13}(1520)$ plays the second important role. 
Its effect is most spectacular around $90^\circ$.
On the contrary, the $F_{15}(1680)$ has important contributions far from $90^\circ$.

Finally, two new resonances seem to be needed to reproduce the data. The new $S_{11}$ resonance 
shows up especially at backward angles and to a less extent at forward angles, with the spanned energy
range located between 1.7 and 1.85 GeV.
Contributions from the new $D_{15}$ resonance are limited to forward angles and high energies.

This section, devoted to the observables of the process
$\gamma p~\to~\eta p$, in the energy range $W~~\lesssim$~2 GeV,
leads to the conclusion that within our approach, the reaction
mechanism is dominant by five known and two new nucleon resonances: 
$S_{11}(1535)$, new $S_{11}$, and to a less extent $D_{15}$, intervene significantly in both observables,
while the $S_{11}(1650)$, and $P_{13}(1720)$ have impact on the cross section, while the 
$D_{13}(1520)$ and $F_{15}(1680)$ play important roles in polarized beam asymmetry.

\subsection{Observables for $\pi^-p\rightarrow \eta n$}\label{Sec:MB}

\begin{figure*}[ht!]

\includegraphics[bb=10 20 470 270 ,scale=1.15]{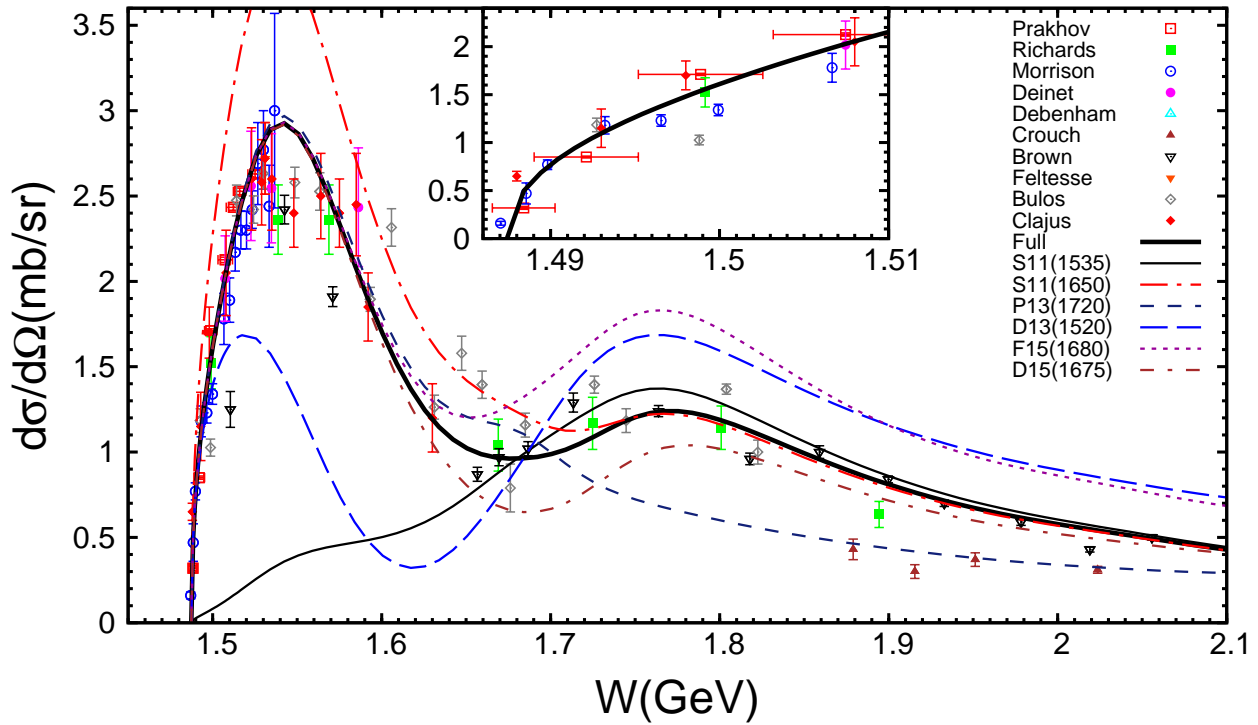}\
\caption{(Color online) Total cross section for $\pi^-p\rightarrow
\eta n$ as a function of $W$.  The curves are: full model (thick full); 
turning off: $S_{11}(1535)$ (thin full), $S_{11}(1650)$ (long dash-dotted),
$P_{13}(1720)$(dashed),  $D_{13}(1520)$(long dashed),
$F_{15}(1680)$(dotted) and $D_{15}(1675)$(short dash-dotted)
resonances.  Data are from Prakhov~\cite{Prakhov2005},
Richards~\cite{Richards1970},
Morrison~\cite{Morrison1999},Deinet~\cite{Deinet1969},
Debenham~\cite{Debenham1975}, Crouch~\cite{Crouch1980},
Brown~\cite{Brown1979}, Feltesse~\cite{Feltesse1975},
Bulos~\cite{Bulos1969,Bulos1964} and Clajus~\cite{Clajus1992}. The
subfigure, inside the box, is for the near threshold energy range. 
\label{Fig:ppent}}
\end{figure*}

We start with presenting the total cross section (Fig.~\ref{Fig:ppent}).
As in the photoproduction case, the $S_{11}(1535)$ brings in the most dominant contribution. 
The contribution of the second $S_{11}$ resonance, $S_{11}(1650)$, turns out to be important, 
though in a restricted part of the phase space. Its vanishing contribution close to 1.7 GeV, 
is compensated by the ape seance of
contributions from the $F_{15}(1680)$. The first $S_{11}$ has a constructive contribution, while that of the second
one and the $F_{15}(1680)$ are destructive. 
The $D_{13}(1520)$ plays the second important role. The peculiar effect of this resonance
could be attributed to strong interferences with other partial waves, which starts as constructive
before turning, around 1.7 GeV, to destructive behavior.
As found in Refs.~\cite{Zhong2007,He2008},
the second peak is from the contribution of $n=2$ shell resonances, and the result in this
work endorses that the $P_{13}(1720)$ accounts for that peak. 
Finally, contributions from the $D_{15}(1675)$ show up around 1.7 GeV.

%
%

Here, we would like to come back to the recent data released by Prakhov {\it et al.}~\cite{Prakhov2005}
in relation to our discussion in Sec.~\ref{Sec:Fit} and Eq.~(\ref{eq:chi2}).
In Fig.~\ref{Fig:ppent}, we have made a zoom, in the box at the top of that Figure, on the
energy region $W \le$ 1.51 GeV. The spread of the energy bin ($\Delta E_{ex}$ in Eq.~(\ref{eq:chi2})) compared
to the cross section uncertainties, and the fact that the cross section rises very rapidly with energy, 
explain the difficulties in fitting the lowest energy data (Fig.~\ref{Fig:ppent}). In that box,
it can be seen that our full model crosses the energy bands, but not always the experimental values
for cross section. 
Actually, using the standard definition of $\chi^2$, those data points give very large contributions and 
render the fitting procedure somewhat problematic. 
That undesirable behavior can be attenuated by embodying the energy bin in the $\chi^2$ determination, as 
in Eq.~(\ref{eq:chi2}).

\begin{figure}[ht!]
 \includegraphics[bb=180 60 390 590 ,scale=.98]{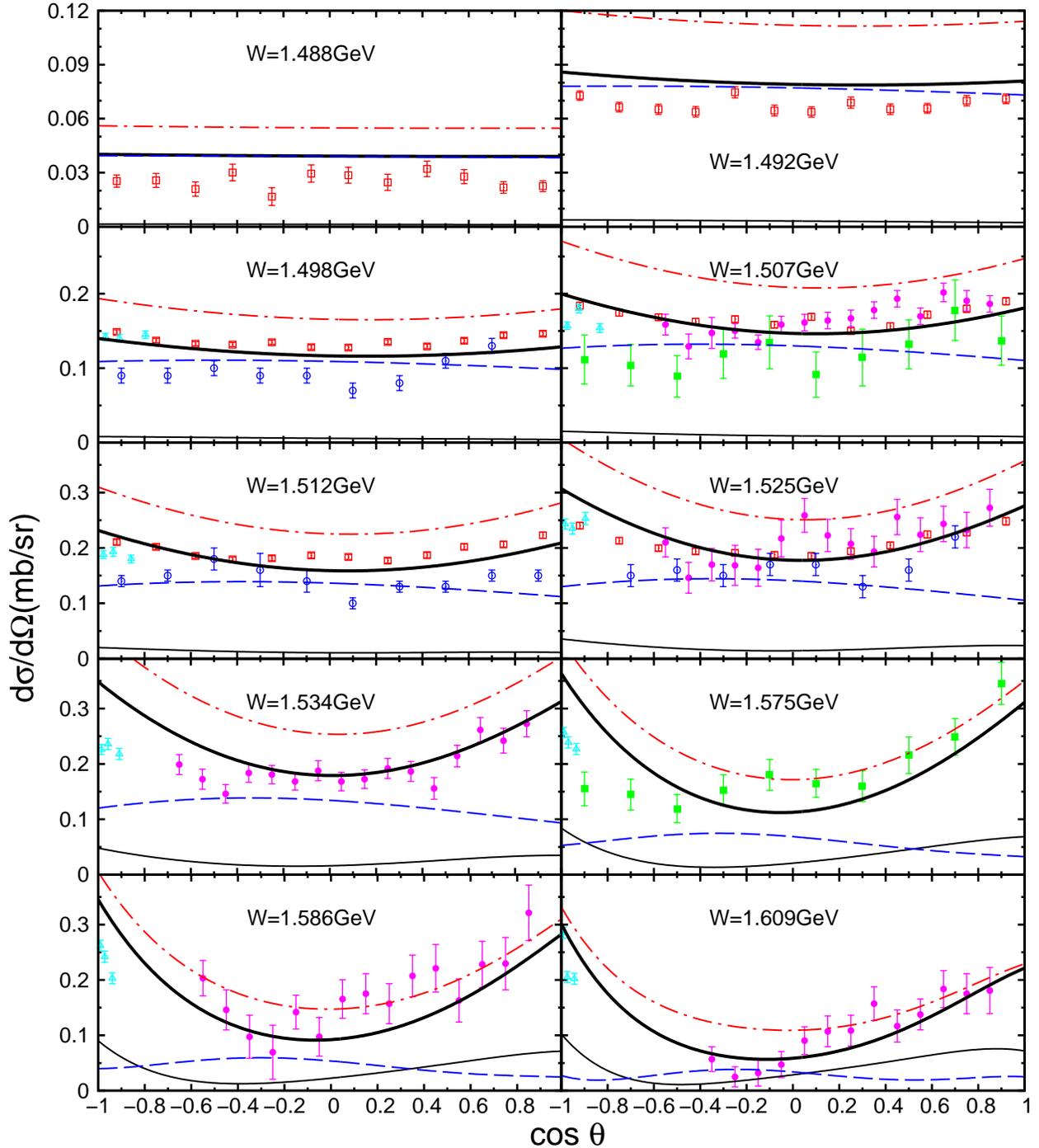}\
\caption{(Color online) Differential cross section for $\pi^-p\rightarrow \eta n$ as a function
of $\cos\theta$ at lower energies. The curves are: full model(thick full); turning off 
$S_{11}(1535)$ (thin full), $S_{11}(1650)$ (dash-dotted), and $D_{13}(1520)$ (dashed). 
Data as Fig.~\ref{Fig:ppent}.
\label{Fig:ppen1}}
\end{figure}

\begin{figure}[ht!]
\includegraphics[bb=170 60 390 480 ,scale=.98]{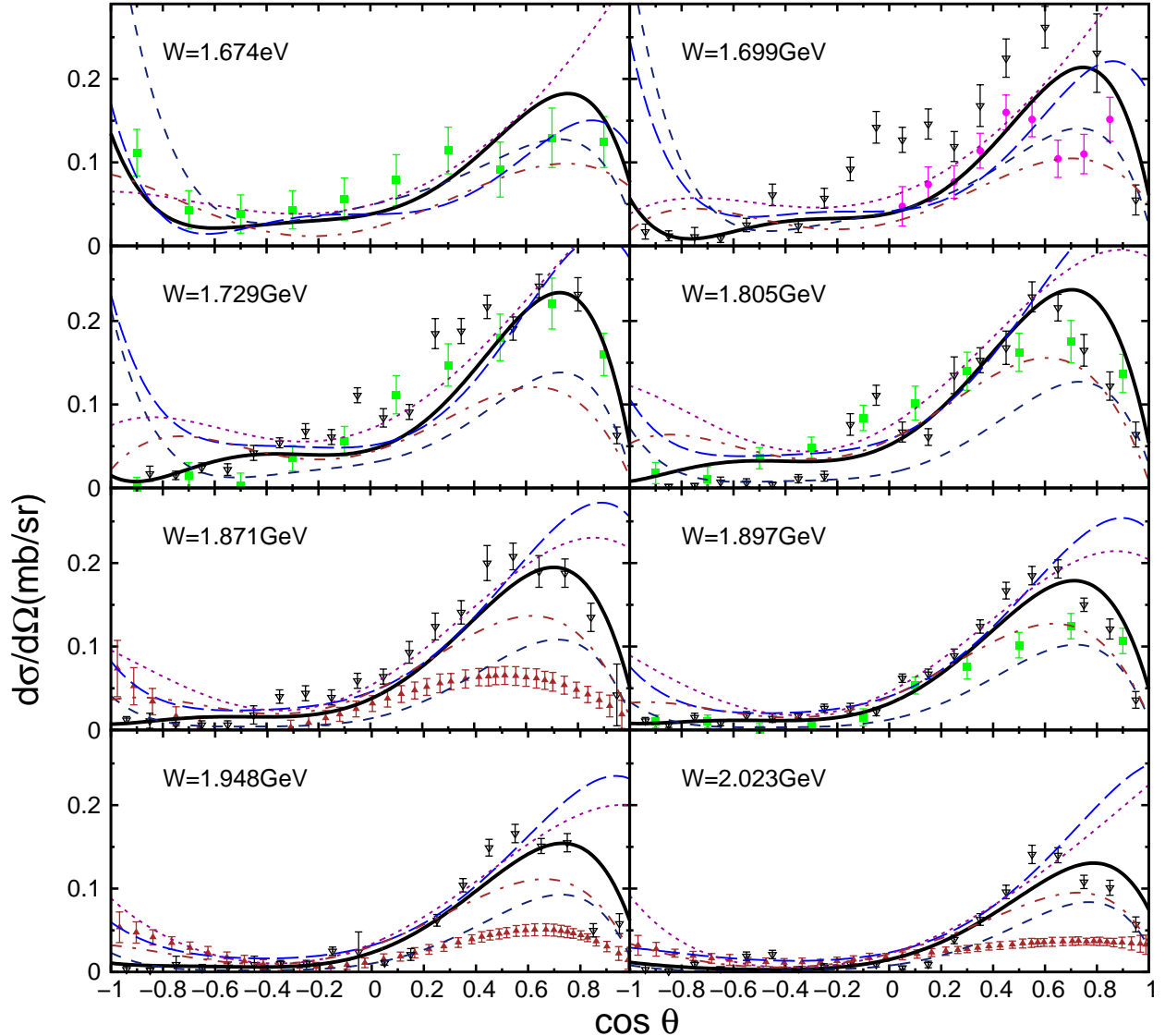}
\caption{(Color online) Same Fig.~\ref{Fig:ppen1}as  at higher energies. 
The curves are: full model (thick full) and turning off: $P_{13}(1720)$ (dashed), $D_{15}(1675)$ (short dash-dotted),
$D_{13}(1520)$ (long dashed), and $F_{15}(1680)$ (thin dashed). 
\label{Fig:ppen2}}
\end{figure}

Now, we move to differential cross section, depicted in Figs.~\ref{Fig:ppen1} and~\ref{Fig:ppen2}, 
and examine the reaction mechanism ingredients. 

The full model reproduces the data reasonably well, with less success in the close threshold region, as  
discussed above.
The highly dominant role of the $S_{11}(1535)$ is present in the whole energy range investigated here, from
threshold up to about 1.6 GeV (Fig.~\ref{Fig:ppen1}). 
Above that energy, as seen in the total cross section plot,
the effect of that resonance becomes rapidly vanishing with energy.
In the low-energy region, $W \lesssim$ 1.6 GeV, we show also results with the $S_{11}(1650)$ or $D_{13}(1520)$
turned off. In both cases the effects are significant, and the latter resonance generates the correct curvature,
required by the data.
%
%

%
%

Figure~\ref{Fig:ppen2} shows results for the higher energy region, where the data are less consistent 
with each other than in the lower energy region. 
Obviously the data by Crouch {\it et al.}~\cite{Crouch1980} can not be simultaneously fitted
with those by Brown {\it et al.}~\cite{Brown1979}. Here, following Ref.~\cite{Durand2008}, we choose the 
latter in the fitting procedure. In going from the low energy to the high energy region the shape
of the theoretical results change: a structure at forward angles appear, while the slope in the backward hemisphere
gets more and more attenuated with energy increase.

The full model reproduces well enough this heterogeneous data base.
By switching off the resonances one by one we show in that Figure, the most significant effects are due to
$P_{13}(1720)$, $D_{13}(1520)$, $F_{15}(1680)$, and $D_{15}(1675)$, which appear mainly at forward hemisphere and to a
less extent in the most backward angles.
The $D_{13}(1520)$ produces the correct curvature at the most forward angles.
Roughly in the same angular region, the $F_{15}(1680)$ plays significant role. Both of those resonances
show destructive contributions. Constructive effects are due to the $P_{13}(1720)$ and $D_{15}(1675)$, with
comparable strengths in the whole forward angle region.

This section, focused on the observables of the process $\pi^-p\rightarrow \eta n$, in the energy range 
$W~\lesssim$~2 GeV, leads to the conclusion that within our approach, the reaction mechanism is dominated by 
six known resonances, $S_{11}(1535)$, $S_{11}(1650)$, $P_{13}(1720)$, $D_{13}(1520)$, $D_{15}(1675)$, and 
$F_{15}(1680)$.

%
%
\subsection{Helicity amplitudes and partial decay widths}\label{Sec:Heli}

After fitting the observables, the helicity amplitudes and the partial decay
widths for $N^*~\to~\eta N$ or $\pi N$ can be calculated within a given model without 
adjustable parameters. 
Results corresponding to our full model are presented in Table~\ref{Tab:Amplitudes} for all
$n$~=1 and 2 shell resonances generated by the quark model including the so-called ``missing'' ones.

\begin{table*}[!ht]

\caption{Helicity amplitudes and decay widths for resonances, with
$\Gamma_{\eta(\pi) N}^{PDG}=\Gamma_{tot}\cdot Br_{\eta(\pi) N}$ in PDG
~\cite{Yao2006}. Here $\sigma$ is the
sign for $\pi N\to~\eta N$ as in Ref.~\cite{Koniuk1980}.}
\renewcommand\tabcolsep{0.16cm}
\begin{tabular}{ l|  c lr r|r r|rr }  \hline\hline
Resonances     & $A_{1/2}$   & $A_{1/2}^{PDG}$  &   $A_{3/2}$  &
$A_{3/2}^{PDG}$  & $\sigma\sqrt{\Gamma_{\eta N}}$ &
$(\sigma)\sqrt{\Gamma_{\eta N}^{PDG}}$ & $\sqrt{\Gamma_{\pi N}}$ &
$\sqrt{\Gamma_{\pi N}^{PDG}}$ \\ \hline
$S_{11}$(1535) & 73& 90 $\pm$ 30   &   &                   &   7.18&     $ 8.87^{+ 1.37}_{-1.37}$        &    6.78 &     $  8.22^{+  1.59}_{-1.60}$   \\
$S_{11}$(1650) & 66& 53 $\pm$ 16   &   &                   &  -2.42&     $ 1.95^{+ 0.94}_{-1.57}$        &    8.85 &     $ 11.31^{+  1.95}_{-1.98}$  \\
$P_{11}$(1440) &-23&-65 $\pm$ 4    &   &                   &  -2.42&                                     &   17.16 &     $ 13.96^{+  4.41}_{ 3.48}$   \\
$P_{11}$(1710) &-53&  9 $\pm$ 22   &   &                   &  -1.05&     $ 2.49^{+ 1.75}_{-0.88}$        &    4.12 &     $  3.87^{+  3.20}_{-1.64}$  \\
$P_{11}$       & 18&               &   &                   &  -2.79&                                     &    6.59 &                                 \\
$P_{11}$       &  3&               &   &                   &  -1.20&                                     &    4.51 &     $  5.34^{+  2.16}_{-2.16} $    \\
$P_{13}$(1720) &177& 18 $\pm$ 30   &-69&  -19 $\pm$ 20     &   2.91&     $ 2.83^{+ 1.04}_{-0.71}$        &   20.15 &     $  5.48^{+  2.27}_{-1.60}$  \\
$P_{13}$(1900) & 30&               &  2&                   &  -1.33&     $ 8.35^{+ 2.11}_{-2.20}$        &   11.02 &     $ 11.38^{+  2.20}_{-2.21}$  \\
$P_{13}$       & 28&               &  0&                   &   2.44&                                     &    3.06 &                                 \\
$P_{13}$       & 12&               &  2&                   &   0.03&                                     &    5.54 &                                  \\
$P_{13}$       & -3&               &  3&                   &  -1.01&                                     &    3.12 &                                 \\
$D_{13}$(1520) & -7& -24 $\pm$  9  &158&  166 $\pm\ $5     &   0.44&     $ 0.51^{+ 0.07}_{-0.06}$        &   14.77 &     $  8.31^{+  0.71}_{-0.53}$   \\
$D_{13}$(1700) & -4& -18 $\pm$ 13  &  4&   -2 $\pm$ 24     &  -0.81&     $ 0.00^{+ 1.22}_{-0.00}$        &    4.92 &     $  3.16^{+  1.58}_{-1.58}$  \\
$D_{15}$(1675) & -6&  19 $\pm$  8  & -8&   15 $\pm\ $9     &  -2.50&     $ 0.00^{+ 1.28}_{-0.00}$        &    7.59 &     $  7.75^{+  0.87}_{-1.00}$   \\
$F_{15}$(1680) & 24& -15 $\pm$  6  &136&  133 $\pm$ 12     &   0.58&     $ 0.00^{+ 1.18}_{-0.00}$        &   13.71 &     $  9.37^{+  0.53}_{-0.54}$    \\
$F_{15}$       & -9&               &  4&                   &   0.97&                                     &    0.35 &                                  \\
$F_{15}$(2000) & -1&               & 10&                   &  -0.47&                                     &    3.60 &     $  4.00^{+  6.20}_{-2.18}$    \\
$F_{17}$(1990) &  5&   1           &  6&  4                &  -1.55&     $ 0.00^{+ 2.17}_{-0.00}$        &    6.84 &     $  4.58^{+  1.55}_{-1.55}$
\\\hline \hline\end{tabular} \label{Tab:Amplitudes}
\end{table*}

The helicity amplitudes  are in line with results from other similar approaches 
(see Tables I and II in Ref.~\cite{Capstick2000}).
Except the $S_{11}(1535)$, the decay widths for $\pi N$ are much larger than those for the $\eta N$ case. 

For the dominant known resonances, $S_{11}$(1535) and $S_{11}$(1650), the helicity amplitudes and decay widths 
for both decay channels are in good agreement with the PDG values.
That is also the case for the $A_{3/2}$ and decay widths of both $D_{13}$(1520) and $F_{15}$(1680). 
For the latter resonance, the $A_{1/2}$ has the right magnitude, but opposite sign with respect to the PDG value. 
However, for that resonance $A_{3/2}$ being much larger than $A_{1/2}$, the effect of
the latter amplitude is not significant enough in computing the observables. 
The helicity amplitudes, as well as $\pi N$ decay width for $P_{13}$(1720) deviate significantly
from their PDG values, as it is also the case in other relevant approaches (see Table II in 
Ref.~\cite{Capstick2000}). 
Those large values produced by our model forced us to treat the symmetry breaking coefficient for $P_{13}$(1720) 
as a free parameter (Table~\ref{Tab: Para}), in order to suppress its otherwise too large contribution. 
As much as the other known resonances are concerned, we get results compatible with the PDG values for 
$D_{13}$(1700) and $F_{17}$(1990), and to a less extent for $D_{15}$(1675). 
Finally, we put forward predictions for the missing resonances, for which we find rather small amplitudes, 
explaining the negligible roles played by them in our model.

\section{Summary and conclusion}\label{Sec:Conclu}

In the present work we have presented a unified description of the processes
$\gamma p\rightarrow \eta p$ and $\pi^-p\rightarrow \eta n$ within a chiral
constituent quark approach, extending our previous investigation of the
photoproduction channel to the  $\pi^-p$ initial state.
Our approach embodies the breaking of the $SU(6)\otimes O(3)$ symmetry,
{\it via} one-gluon-exchange mechanism. The generated configuration mixing is
characterised by mixing angles, which we have determined without specific
free parameters. Moreover, the present quark approach
is used to derive photoexcitation helicity amplitudes and partial decay
widths of the nucleon resonances to $\pi N$ and $\eta N$ final states.

Our study is focused on the reaction mechanisms of the considered reactions
in the energy range from threshold up to the centre-of-mass energy $W~\approx$~2 GeV,
where data for both reactions are available. Accordingly, the nucleon resonances
taken into account are explicitly dealt with for $n \le$~2 harmonic-oscillator shells.
Within that frame-work, we have Successfully fitted close to 1800 data points with 21 adjustable parameters, 
with 9 of them related to the 3 new resonances. With such a rather small number of
free parameters, we have  investigated possible roles played in those reaction mechanisms by 
12 known resonances, 6 the so-called missing resonances, and 3 new ones.
 
The combined fit of the known baryon spectrum and the $\gamma p\rightarrow \eta p$
measured observables, allowed us to i) extract the masses of missing resonances generated
by the used formalism, ii) extract the mixing angles between relevant configurations, which came
out~\cite{He2008a} in agreement with Isgur-Karl pioneer work, iii) determine the parameters of the 3 new
resonances, compatible with other findings.

The reaction mechanism for the process $\gamma p\rightarrow \eta p$ is found, as expected,
being dominated by the $S_{11}(1535)$, with significant contributions from four additional
known resonances ($S_{11}(1650)$, $D_{13}(1520)$, $F_{15}(1680)$, and $P_{13}(1720)$), and
from a new resonance $S_{11}(1717)$. The importance of those five known resonances is corroborated
by the calculated photo-excitation helicity amplitudes and final-state $\eta p$ branching ratios.

For the photoproduction channel, the new $S_{11}$ resonance turns out to be essential in reproducing the data, for which the 
extracted Wigner mass and  width come out consistent with
the values in Refs.~\cite{Saghai2001,Li1996,Batinic1997,Julia-Diaz2006},
but the mass is lower, by about 100 to 200 MeV, than findings by other
authors~\cite{Capstick1994,Giannini2001,Chen2003,Chiang2003,Tryasuchev2004}.
The most natural explanation would be that it is the first $S_{11}$
state in the $n=3$ shell. However its low mass could indicate a multiquark component, such as, 
a quasi-bound kaon-hyperon~\cite{Li1996} or  pentaquark configuration~\cite{Zou2007}.
In Ref.~\cite{Kuznetsov2008},the authors propose a N*(1685) from the reanalysis of the GRAAL 
beam asymmetry data, but their results do not support a $S$-partial wave resonance.

Cutkosky~\cite{Cutkosky1979} reported a $D_{13}$ resonances
at (1880$\pm$100) MeV with (180$\pm$60) MeV widths.
Recent investigation find $D_{13}(1875)$ a state coupling
strongly to kaon-hyperon channels but not to the $\eta N$
channel~\cite{Mart2000,Anisovich2005}.
In this work, we also find for the new $D_{13}$ resonance the variation of
$\chi^2$ is small compared with other resonances. 
Interestingly, we find large effect from a $D_{15}$ state around
2090 GeV with a Wigner width of 330 MeV. It is very similar to the
$N(2070)D_{15}$ reported in Refs.~\cite{Crede2005,Sarantsev2005,
Anisovich2005}. It can be explained as the first $D_{15}$ state in $n=3$
shell~\cite{Crede2005}.

We come back to the known $P$- and $D$-wave resonances, in the mass energy region around 1.71 GeV,  
and hence the most important energy region to study the $n=2$ shell resonances, in both processes
investigated here. 
In the literature, different conclusion have been reported on
the relative weight of the relevant resonances  in the reaction mechanism. 
In Ref.~\cite{Nakayama2008}, the inclusion of the $P_{13}(1720)$ resonance does not improve significantly the 
description of the data for the photoproduction, while this resonance considerably improves the fit quality of
the hadronic $\pi^-p\rightarrow \eta n$ reaction at higher energies,  and the small bump near W = 1.7 GeV in 
the spin-1/2 resonance contribution is attributed to the P11(1710) resonance.
In Ref.~\cite{Shklyar2004}, this latter resonance together with the background contributions dominate the 
$\pi^-p\rightarrow \eta n$ reaction in the $n=2$ shell energy area, developing a peak in the total cross section 
around 1.7 GeV.
As in Ref.\cite{Shyam2008}, our results endorse the n=2 shell
resonances $P_{13}(1720)$, not the $P_{11}(1710)$, providing the most
significant contribution in both $\eta$ production processes. 
A recent quark model approach~\cite{Zhong2007} found important effects due to the $P_{11}(1720)$ in the 
$\pi^-p\rightarrow \eta n$, but did not consider the $SU(6)\otimes O(3)$-symmetry breaking.
In the present work, the crucial character of this latter mechanism  attenuate the role
attributed to $P_{11}(1720)$.
However, our results do not allow us to reach firm conclusions about the role of the $P$-wave resonances. 
The origin of this situation can be traced back to the discrepancies between the calculated photo-excitation amplitudes 
for the $P_{13}(1720)$  and those reported in PDG.
Such discrepancies are also found in
other constituent quark model calculations~\cite{Capstick2000}. This may
indicated that the $P_{13}(1720)$ has a more complicated structure than the
simple three quark picture.
In Ref.~\cite{Shyam2008}, the $D_{13}(1700)$ resonance gives the largest contribution to the cross section in the energy
region of $W$ = 1.7-2.0 GeV, but we do not find such contributions from the $D_{13}(1700)$.

To summarize our findings with respect to the $\pi^-p\rightarrow \eta n$, we emphasize that,
within our approach besides the (by far) dominant $S_{11}(1535)$, significant contributions
from the same four known resonances ($S_{11}(1650)$, $D_{13}(1520)$, $F_{15}(1680)$, and $P_{13}(1720)$)
are found, as in the photoproduction case. For the strong channel, another resonance turns out to relevant, namely
$D_{15}(1675)$.

From theoretical point of view, the next steps are i) perform a comprehensive extended coupled-channels study 
~\cite{Durand2009b}, 
ii) extend the present formalism to higher $n$ shells in order to investigate all photoproduction data
up to $W~\approx$ 2.6 GeV, embodying all PDG one to four star resonances~\cite{He2009}.
In the experimental sector, the most needed data concern the $\pi^-p\rightarrow \eta n$ channel. Double polarization
observables for $\gamma p\rightarrow \eta p$ planned to be measured at JLab~\cite{Crede2007}
will certainly improve our understanding of the underlying reaction mechanisms.
%
%

\section*{Acknowledgements}

We are deeply grateful to Johan Durand for providing the data base for $\pi^-p\rightarrow \eta n$.

\end{document}